\documentclass[prl,twocolumn,showpacs,nobibnotes]{revtex4}

\usepackage{bm}
\usepackage[dvips,final]{graphics}

\begin{document}
 
\title{Time-dependent density-matrix renormalization-group: 
A systematic method for the study of quantum many-body systems
out-of-equilibrium}
 
\author{M. A. Cazalilla}
\affiliation{The Abdus Salam ICTP, Strada Costiera 11, 34014 Trieste, Italy}
\affiliation{Donostia International Physics Center (DIPC), 
P. Manuel de Lardizabal 4, 20018 Donostia, Spain}

\author{J. B. Marston}
\affiliation{Department of Physics, Brown University,
Providence, RI 02912-1843, USA}

\date{\today}
 
\begin{abstract}
The density-matrix renormalization-group (DMRG) algorithm 
is extended to treat time-dependent problems. 
The method provides a systematic and robust 
tool to explore out-of-equilibrium phenomena in quantum many-body systems. 
We illustrate the method by showing that attractive interactions 
enhance the tunneling current between two Luttinger liquids, whereas
repulsive interactions suppress it, in qualitative agreement with
analytical predictions.  Enhancement of the transport current 
through a quantum dot in the Kondo regime is also exhibited. 
\end{abstract}
 
 
\maketitle

The study of out-of-equilibrium phenomena in quantum 
many-body systems has mainly focused on steady state properties, 
such as the conductance of quantum dot systems\cite{qdotsth,qdotsexp} 
and of point contacts between two Luttinger liquids\cite{Kane92,Yao99}.  More 
recently, attention has been paid to time-dependent phenomena 
in these systems\cite{Wingreen93,Schiller9600,Ng96, Nordlander99,
Plihal99,Kaminski99,Kouwenhoven97,Cuniberti98,Schmeltzer00,Talyanskii97}. 
One goal of the work has been to identify the 
hierarchy of time scales which arise when quantum-many body systems 
are driven out of equilibrium by a variety of time-dependent perturbations.  
This information is relevant for experiments involving potentially
important technological applications of quantum dots 
and junctions built from carbon nanotubes, and quantum 
wires which behave as Luttinger liquids (LL)\cite{Yao99,LL}. 
However, many studies have employed either uncontrolled approximations, 
or exact solutions at special values of the model parameters which are 
rather remote from experimentally accessible systems. 

By contrast, the density-matrix renormalization-group 
(DMRG) algorithm\cite{White929398,Peschel98} 
has been successfully employed to calculate 
many static and frequency-dependent properties of quasi one-dimensional
quantum many-body systems\cite{Peschel98}.  The numerical method 
does not depend on whether the system is integrable or not, 
and in many cases it provides essentially exact results 
as errors induced by truncating the Hilbert space are controlled in
a systematic way.  

We employ the so-called infinite-size 
DMRG\cite{White929398,Peschel98} as it treats the center of the chain,
the region where we focus our attention, most accurately.
Our starting point is a quantum Hamiltonian representing a chain of 
$L$ sites.  If we denote by $D$ the dimension of the Hilbert space on 
each site, then the dimension of the Hilbert space
${\cal H}$ is $D^{L}$.  The DMRG algorithm organizes the Hilbert space
into blocks, ${\rm B_{L}}\,\bullet\,\bullet\,{\rm B_{R}}$, 
where $\bullet\,\bullet$ represents the Hilbert space of the two central 
sites (of dimension $D$) and ${\rm B_{L}}$, ${\rm B_{\rm R}}$ are respectively 
the blocks representing the Hilbert space of the remaining left and right sites, 
each of dimension $D^{L/2 -1}$.  An initial small chain of
$L = 4$ sites is first enlarged to 6 sites by cutting it in half and inserting 
two sites at the center.  As this process is repeated, the Hilbert space 
grows exponentially and eventually exceeds a limit, $(D * M)^2$, beyond
which it is truncated.  As the calculation may be systematically 
improved by increasing the size of the retained block Hilbert space, $M$, up to 
limits imposed by computer memory and speed, it is straightforward to check
for convergence in the observables.  The end result of the algorithm is an 
approximation to the exact many-body ground state 
$|\Psi^{\rm trunc}_{\mathrm o}\rangle$, the ground state energy $E_0$,
and a truncated representation of the Hamiltonian, $H_{\rm trunc}$. 

\begin{figure}
\resizebox{8cm}{!}{\includegraphics{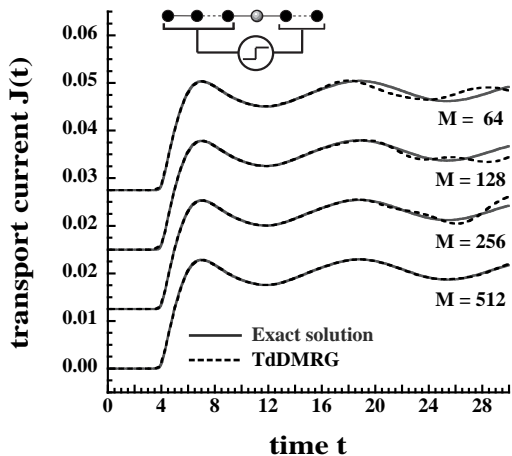}}
\caption{Transport current $J(t)$ for the spinless quantum dot system 
defined by Eq.(\ref{eq2}). The plots (which are offset in the vertical
direction for clarity) illustrate systematic convergence towards the
exact independent-particle solution with increasing block dimension 
$M$. Time is measured in units of $\tau_{w} \equiv \hbar/w$ and $J(t)$ 
is expressed in units of $e/\tau_{w}$.  The parameters are $t_q = 0.15\: w$,
$\varepsilon_q = -0.25 \: w$, and $L = 64$ sites.  At time 
$t_{\mathrm o} = 4 \, \tau_{w}$ the chemical potential of the leads is 
shifted by a bias of $\delta \mu_{\mathrm o} = -0.25\, w$ within a rise time of
$t_{s} = 0.1\, \tau_{w}$. The period of the ringing oscillations is given 
by the expression reported in Ref.~\onlinecite{Wingreen93}:
$\Delta t_{R} = 2\pi \hbar/| \delta \mu_{R}-\varepsilon_q|
= 4\pi \: \tau_{w}$.}
\label{fig1}
\end{figure}

The extension to time-dependent problems, which we call  
time-dependent DMRG (TdDMRG), takes the infinite-size DMRG algorithm
described above as its starting point.  After the penultimate iteration,
when the chain has $L - 2$ total sites, either a quantum dot paired with
an ordinary site, or two sites linked by a tunneling junction 
(depending on the choice of problem to be studied) are inserted at
the center of the chain, yielding a chain with
a total of $L$ sites.  Finally, a time-dependent 
perturbation, $H'(t)$ is added to $H_{\rm trunc}$, and the resulting  
time-dependent Schr\"odinger equation is then integrated forward in time:
\begin{equation}\label{eq1}
i \hbar \frac{\partial}{\partial t}| \Psi(t) \rangle = \big[ 
(H_{\rm trunc} - E_0) + H'(t) \big] |\Psi(t)\rangle\ .
\end{equation}
The initial state is chosen to be the ground state of the unpertured
truncated Hamiltonian,  $|\Psi(t = 0)\rangle = |\Psi^{\rm trunc}_{\mathrm{o}}\rangle$. 

To illustrate the method, we apply the TdDMRG algorithm to two kinds of systems:
quantum dots and tunneling junctions between two chains 
of one-dimensional interacting fermions. 
We first consider the easier problem of spinless fermions (with small on-site Hilbert space
dimension $D=2$) before turning to a more realistic model involving 
electrons with spin. 

1. {\bf Spinless Fermions}: 
We turn first to the problem of a quantum dot coupled to two leads.  
A similar problem has been studied by Wingreen et al.\cite{Wingreen93}
in the thermodynamic $L \to \infty$ limit.  In terms of the fermion creation and 
annihilation operators $c^{\dagger}_j, c_{j}$, and the number operators
$n_j \equiv c^\dagger_j c_j$, the time-independent part of 
the Hamiltonian for the quantum dot system reads:
\begin{eqnarray}\label{eq2}
H_{\rm qdot} &=& -\frac{w}{2}\:\sum_{j \neq q-1, \: q}\:\left[ 
c^{\dagger}_{j+1} c_{j} + {\rm H.c.} \right] + \varepsilon_q 
\: n_q \nonumber \\
&-& t_q \: \left[ c^{\dagger}_{q} c_{q-1} +  c^{\dagger}_{q+1} 
c_{q}  + {\rm H.c.} \right].
\end{eqnarray}
In this equation $q$ denotes the 
location of the right central site where the quantum dot is located, 
$q = L/2+1$.  Here $t_q$ is the hopping amplitude between the dot and
either lead, and $w > 0$ is the  half band-width of the leads, which 
are half-filled.  The local chemical potential at the quantum dot 
site, $\varepsilon_q$, is generally nonzero.  It may be viewed as the
energy of a localized level.   

The time-dependent perturbation is a differential bias applied between
the left and right leads:  
\begin{equation}\label{eq6}
H'(t) = -\delta \mu_R(t) N_{\rm R} - \delta \mu_L(t) N_{\rm L}\,.
\end{equation}
We typically set $\delta \mu_L(t) = -\delta \mu_R (t) = \delta \mu_{\rm o} \: 
\theta(t-t_{\rm o})$, where $\theta(t)$ is  a smoothed step function with rise 
time $t_s$: $\theta(t) = 1/[\exp(-t/t_{s})+1]$. 
Operators $N_{\rm R} = \sum_{j= q+1}^{L} n_j$ and $N_{\rm L}$ (which has 
a similar expression) count the number of fermions in the right and left leads. 
  
It is convenient to gauge-transform the bias into a local, but time-dependent, 
hopping amplitude by means of the transformation
$|\bar{\Psi}(t)\rangle  = 
\ e^{i \Phi(t) \left[ N_{\rm R} - N_{\rm L} \right]/\hbar} \: 
|\Psi(t)\rangle$, where $\Phi(t) = \int_{-\infty}^{t} 
\delta \mu_L(t^\prime)~ dt^\prime$.  Thus the real-valued hopping 
amplitude $t_q$ is replaced by a complex-valued one: 
$\bar{t}_q(t) = t_q \: e^{i \Phi(t)/\hbar}$,  
and the differential bias, which acts on all the sites in the blocks, 
is absorbed into an operator that acts only locally on the 
central region of the chain where the infinite-size DMRG 
is most accurate.   

The expectation values of the currents along the central links 
may be evaluated during the course of the wavefunction's 
time-evolution\cite{RungeKutta} by calculating: 
\begin{equation}\label{eq7}
J_{q,q-1}(t) = -\frac{2e}{\hbar} {\rm Re} \Big\{ i~ 
\bar{t}_q(t)\: \langle \bar{\Psi}(t) | c^{\dagger}_{q} c_{q-1} | 
\bar{\Psi}(t) \rangle \Big\}\ .
\end{equation}
Following Refs.~\onlinecite{Wingreen93} and \onlinecite{Plihal99} 
we introduce an average transport current defined as: 
$J(t) = \frac{1}{2} [J_{q,q-1}(t)  + J_{q+1,q}(t)]$. 
The results of our calculation are shown in Fig.~\ref{fig1}. 
The exact solution for non-interacting systems is obtained by integrating the
equations of motion for the quantities $\gamma_{ij} = \langle c^{\dagger}_i(t)
c_j(t)\rangle$ forward in time for systems of the same finite size $L$; 
therefore the results can be compared directly with the TdDMRG calculations.

Another problem of recent interest is transport through a point contact 
or junction between two Luttinger liquids.  As the DMRG algorithm 
works equally well for interacting leads, we may turn on 
nearest-neighbor interactions $V n_j n_{j+1}$ between 
the spinless fermions (again $q = L/2 + 1$):  
\begin{eqnarray}\label{eq3}
H_{\rm junct} &=& -\frac{w}{2}\: \sum_{j\neq q-1}\:\left[ 
c^{\dagger}_{j+1} c_{j} + {\rm H.c.} \right] \nonumber \\ 
&+& V \: \sum_{j \neq q-1} (n_{j+1} - \frac{1}{2})  
(n_{j} -\frac{1}{2}) \nonumber \\
&-& t_q \: \left[ c^{\dagger}_{q} c_{q-1}  + 
{\rm H.c.} \right]. 
\end{eqnarray}
At half-filling the leads are metallic for $|V| < w$ and exhibit LL 
behavior\cite{Haldane82}; otherwise a charge-density wave (CDW) 
forms and an insulating gap opens up.
Kane and Fisher\cite{Kane92} studied this model in the continuum
limit for $V < w$.  Within perturbative RG and 
bosonization they concluded that intralead
interactions strongly affect the transmittance of the junction.  For attractive 
interactions ($V < 0$), the transmittance is enhanced whereas for repulsive 
potentials ($V > 0$) it is suppressed.   
 
\begin{figure}
\resizebox{8cm}{!}{\includegraphics{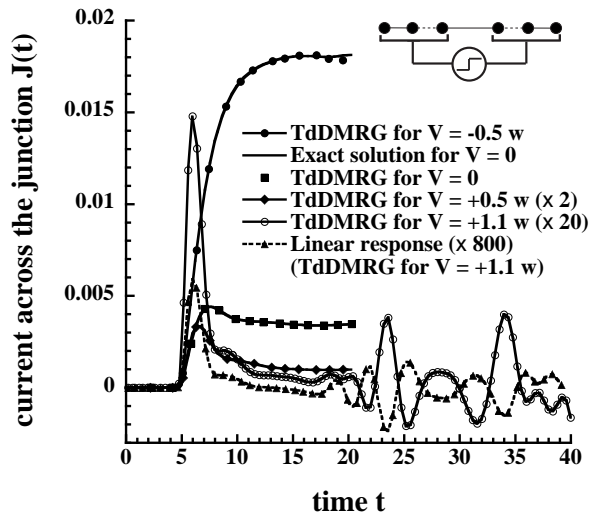}}
\caption{Current across a junction between two systems of one-dimensional
interacting electrons in response to a step bias applied between the two leads. 
Interactions strongly affect the currents, in qualitative agreement 
with the predictions of Ref.~\onlinecite{Kane92}. 
The junction consists of a weakened link with $t_q = 0.125~ w$
lying at the center of a chain of $L= 64$ sites.  At time 
$t_{\mathrm o} = 5~ \tau_{w}$ the bias is shifted by 
$\delta \mu_{\mathrm o} = 0.0625~ w$ in a rise time of 
$t_{\rm s} = 0.1~ \tau_{w}$.  The label ``linear response''
refers to the application of a much smaller bias 
$\delta \mu_{\mathrm o} = 6.25 \times 10^{-4}~ w$.  
Lines are for the more accurate $M = 512$ truncation 
(or, in the non-interacting $V=0$ case, 
depict the exact solution) while data points are for $M = 256$.  
For clarity, currents have been multiplied by factors of 
2, 20, and 800 as indicated.  
Note the good convergence with increasing blocksize $M$, and 
with the exact result in the non-interacting case. 
The units of current and time are the same as in Fig.~\ref{fig1}. 
The reflection time $t_{\rm refl} = 64 \tau_w$ is longer
than the time period shown.}
\label{fig2}
\end{figure} 

Results for the two systems of spinless fermions are shown in 
Figs.~\ref{fig1} and \ref{fig2}.  In both cases the short time behavior of
the current compares well with the available 
exact independent-particle results, 
even for rather small block Hilbert space dimension $M$ 
(see also Fig.~\ref{fig3} below).  This may seem surprising, as only 
the ground state was targeted in the DMRG portion of the algorithm.  
However it is important to note that the DMRG algorithm is not a true
renormalization-group method as the reduced density-matrix, rather than 
energetics, determines which states are projected-out of the Hilbert space.
Over short times the perturbation can only create a limited number of
excitations about the ground state, so the time-evolved state remains
closely connected with the ground state and can be accurately represented 
with a superblock of relatively small Hilbert space dimension.  
Fig.~\ref{fig1} also illustrates the systematic convergence with increasing 
blocksize $M$.  By retaining more Hilbert space we are able to extend the 
accurate solution forward in time.  Nevertheless even short-time behavior often 
suffices to reveal interesting physics.  For example the initial behavior 
of the junction current depicted in Fig.~\ref{fig2} is consistent with 
the prediction of Kane and Fisher\cite{Kane92}:  An attractive interaction  
enhances the current $J(t)$ whereas a repulsive interaction suppresses it. 
It is interesting to observe that in the insulating CDW regime 
$V > w$ current oscillations develop after the initial transient.  
We emphasize that the TdDMRG method enables us to access short-time 
out-of-equilbrium phenomena in an accurate way, 
and thus goes well beyond bosonization.

Several other aspects of the TdDMRG method are worth mentioning.
First, as the numerical method necessarily addresses only systems of
finite size, the current averaged over long times must always vanish, 
regardless of the bias.  This is so because any current pulse created 
by the bias eventually reaches the open ends of the chain within a time 
of order $t_{\rm refl} = L a / v_{F}$,
where $v_{F} = w\: a/\hbar$ is the Fermi velocity and $a$ the lattice spacing.  
At that point the pulse is reflected back, leading to oscillations in the
current.  To be useful, model parameters must be chosen such that the
physics of interest occurs over time scales shorter than $t_{\rm refl}$. 
There is no limit on the size of the applied bias: 
large biases can be applied, driving the system far outside of the 
linear-response regime, though we find that larger biases generally require
larger blocksizes $M$ to maintain the accuracy of the solution.  
Sizeable biases were applied in the above 
two model problems, Figs.~\ref{fig1} and ~\ref{fig2}. 
The non-linear effects of the bias are illustrated by the response in
the $V > w$ regime shown in Fig.~\ref{fig2}, as the form of the oscillations 
depends strongly on the size of the applied bias.  

\begin{figure}
\resizebox{8cm}{!}{\includegraphics{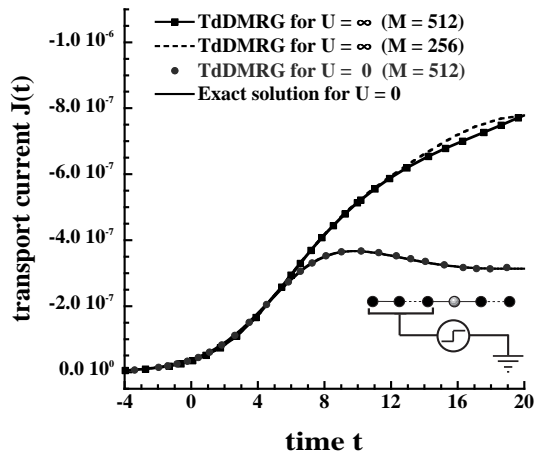}}
\caption{Transport current $J(t)$ for a quantum dot in the Kondo regime 
driven out of equilibrium by a step bias applied to the left lead.  
$J(t)$ is enhanced compared to the non-interacting $U = 0$ case.  
The parameters used in the calculation are $L = 100$, $t_q = w/(4\sqrt{2})$, 
$\varepsilon_q = -0.25 \: w$, and $U = \infty$, with a corresponding Kondo
energy scale of order $k_B T_{K} = 0.003\, w$. The step in the bias is turned 
on at time $t_{\mathrm o} = 4\, \tau_{w}$ within a rise time of 
$t_{s} = 2 \, \tau_{w}$.  The bias reduces the chemical potential of the 
left lead by a small amount, $\delta \mu_{\mathrm o} = -5 \times 10^{-6} w$. 
As the time advances small discrepancies show up between the more accurate
$M=512$ truncation (continuous curve with filled squares) and 
the $M=256$ truncation (dotted curve). The units of current and time are 
the same as in Fig.\ref{fig1}.}
\label{fig3}
\end{figure}

2. {\bf Spinning Electrons}: To illustrate the application of the 
TdDMRG method to a more realistic problem involving spinning electrons, 
we turn now to the study of the one-impurity Anderson model described by 
the Hamiltonian: 
\begin{eqnarray}\label{eq8}
H_{\rm qdot} &=& -\frac{w}{2}\:\sum_{\sigma j \neq q-1,~ q}\:
\left[ c^{\dagger \sigma}_{j+1} c_{j \sigma} + {\rm H.c.} \right] \nonumber \\
&+& \varepsilon_q \: n_q  + U\: n_{\uparrow q} n_{\downarrow q} \nonumber \\
&-&  t_q \sum_{\sigma} \left[ c^{\dagger \sigma}_{q}  
c_{q-1 \sigma} +  c^{\dagger \sigma}_{q+1} c_{q \sigma} + {\rm H.c.} \right].
\end{eqnarray}
Here $\sigma = \uparrow, \downarrow$ labels the spin and 
$n_q \equiv n_{q \uparrow} + n_{q \downarrow}$ is the total number of 
electrons at the dot site.
The one-impurity Anderson model is a minimal model to describe a quantum dot 
with spinning electrons coupled to leads\cite{qdotsth}. 
The model also arises in studies of strongly correlated electrons systems 
in the limit of infinite spatial dimension, in particular within 
dynamical mean field theory (DMFT)\cite{DMFT}.
The method presented here can be adapted to the calculation of 
the imaginary-time Green's functions needed in the DMFT 
calculations\cite{unpub}.   

In this case only the bias of the left lead is shifted, 
$H'(t) = -\delta \mu_L(t) N_{\rm L}$, but again this shift can be 
gauge-transformed into a time-dependent hopping amplitude. 
Results for the initial transport current $J(t)$ are shown 
in Fig.~\ref{fig3}.  Here $\varepsilon_q = -0.25 w$ and the width 
$\Delta = 0.125 w$; thus in the large-U limit the quantum dot is 
in the Kondo regime. For these parameters 
the Kondo scale is estimated to be of order 
$k_B T_{K} = 0.003 w$\cite{Hewson}.  For comparison
we also plot the current in 
the non-interacting $U = 0$ limit, calculated both
exactly within the independent-particle 
picture, and via TdDMRG. The current
for $U = \infty$ shows the expected 
Kondo enhancement\cite{qdotsth, Plihal99}.  
Upon decreasing the hopping amplitude 
between the dot and the leads to $t_q = 0.05 w$
we find by contrast that the current at $U = \infty$ is less than at
$U = 0$.  We attribute this drop to the existence of a Coulomb blockade, 
which now overwhelms the vanishingly 
small Kondo resonance  [$k_B T_{K} = O(10^{-19} w)$].
A more detailed discussion of our results, also in other settings,
will be given elsewhere\cite{unpub}.

In conclusion, we have shown that the  
DMRG algorithm can be extended to treat time-dependent problems. The TdDMRG
method is a systematic tool to study 
the transient behavior of quantum many-body 
systems. The method has been illustrated by applying it to three systems: 
spinless and spinful quantum dots and a junction between two Luttinger liquids. 
Our results agree with previous predictions but do not rely on uncontrolled
approximations.  Reliable results can be obtained, even for systems driven 
far out-of-equilibrium. 

We thank N. Andrei, L. Glazman, K. Hallberg, 
A.~F. Ho, C. Hooley, J. Merino, 
O. Parcollet, J. von Delft, and S. White for useful discussions. 
This work has been supported by the {\it Hezkuntza, Unibertsitate eta Ikerketa 
Saila} of the Government of the Basque Country, and by the US NSF through 
grant no. DMR-9712391.


\begin{thebibliography}{60}
  
\bibitem{qdotsth}
L. I. Glazman and M. E. Raikh, PisUma Zh. \'Eksp. Teor. Fiz. 
{\bf 47}, 378 (1988) [JETP Lett. {\bf 47}, 452 (1988)]; T. K. Ng and 
P. A. Lee, Phys. Rev. Lett. {\bf 61}, 1768 (1988).
 
\bibitem{qdotsexp}
D. Goldhaber-Gordon, J. G{\"o}res, M. A. Kastner,
H. Shtrikman, D. Mahalu, and U. Meirav,
Phys. Rev. Lett. {\bf 81}, 5225 (1998).
  
\bibitem{Kane92}
C. L. Kane and M. P. A. Fisher, Phys. Rev. B {\bf 46}, 15233 (1992).
 
\bibitem{Yao99}
Z. Yao, H. W. Ch. Postma, L. Balents, and C. Dekker, Nature {\bf 402}, 273 (1999).
 
\bibitem{Wingreen93}
N. S. Wingreen, A.-P. Jauho, and Y. Meir, Phys. Rev. B {\bf 48}, 8487 (1993).
  
\bibitem{Schiller9600}
A. Schiller and S. Hershfield, Phys. Rev. Lett. {\bf 77}, 1821 
(1996); Phys. Rev. B {\bf 62} R16271 (2000).
 
\bibitem{Ng96}
T. K. Ng, Phys. Rev. Lett. {\bf 76}, 487 (1996).
 
\bibitem{Nordlander99}
P. Nordlander, M. Pustilnik, Y. Meir, N. S. Wingreen, and D. C. Langreth,  
Phys. Rev. Lett. {\bf 83}, 808 (1999).
  
\bibitem{Plihal99}
M. Plihal, D. C. Langreth, and P. Nordlander, Phys. Rev. B {\bf 61}, R13341 (1999).
 
\bibitem{Kaminski99}
A. Kaminski, Yu. V. Nazarov, and L. I. Glazman, Phys. Rev. Lett. {\bf 83}, 384 (1999). 
 
\bibitem{Kouwenhoven97}
L. P. Kouwenhoven, C. M. Marcus, P. L. McEuen, S. Tarucha, 
K. M. Westervelt, and N. S. Wingreen, ``Transport in Quantum Dots,'' 
in {\it Mesoscopic Electron Transport}, L. L. Sohn, L. P. Kouwenhoven, 
and G. Sch\"on (Eds.) Kluwer (The Netherlands, 1997).
 
\bibitem{Talyanskii97}
V. I. Talyanskii, J. M. Shilton, M. Pepper, C. G. Smith, C. J. B. Ford, 
E. H. Linfield, D. A. Ritchie, and G. A. C. Jones, Phys. Rev. B {\bf 56}, 15180 (1997).
 
\bibitem{Cuniberti98}
G. Cuniberti,  M. Sassetti, and B. Kramer, Phys. Rev. B {\bf 57}, 1515 (1998).
 
\bibitem{Schmeltzer00}
D. Schmeltzer, Phys. Rev. Lett. {\bf 85}, 4132 (2000).
 
\bibitem{LL} 
M. Bockrath, David H. Cobden, Jia Lu, Andrew G. Rinzler,
Richard E. Smalley, Leon Balents, and Paul L. McEuen, 
Nature {\bf 397}, 598 (1999); O. M. Auslaender, A. Yacoby,
R. de Picciotto, K. W. Baldwin, L. N. Pfeiffer, and K. W. West,
Phys. Rev. Lett. {\bf 84}, 1764 (2000).
 
\bibitem{White929398}
S. R. White, Phys. Rev. Lett. {\bf 69}, 2863 (1992); Phys. Rev. B 
{\bf 48}, 10345 (1993); Phys. Rep. {\bf 301}, 187 (1998).
  
\bibitem{Peschel98}
I. Peschel, X. Wang, M. Kaulke, and K. Hallberg (Eds.)
{\it Density-Matrix Renormalization: A New Numerical Method
in Physics}, Springer-Verlag (Berlin, 1999).
 
\bibitem{RungeKutta}
We use a fourth-order Runge-Kutta algorithm to
integrate the Schr\"odinger equation forward in time.
The wavefunction preserves its normalization to better than 
a part in $10^8$ over the course of its time-evolution. 
We also check that the charge is conserved
both globally and at each of the central sites. Furthermore, for spinful
electrons, the spin moments remain extremely small during the time evolution,
as they must.

\bibitem{Haldane82}
F. D. M. Haldane, Phys. Rev. B {\bf 25}, 4925 (1982).
 
\bibitem{DMFT}
A. Georges, G. Kotliar, W. Krauth, and M. J. Rozenberg,  
Rev. Mod. Phys. {\bf 68}, 13 (1996).
 
\bibitem{unpub}
M. A. Cazalilla and J. B. Marston, unpublished.
 
\bibitem{Hewson}
A. C. Hewson, {\it The Kondo Problem to Heavy Fermions} 
Cambridge University Press (Cambridge, 1993), page 182.
 
\end{thebibliography}
\end{document}